**Comment on "Levitation and Self-Organization of Liquid Microdroplets over Dry Heated Substrates"**

In a recent letter [1], Zaitsev et al. report observations of evaporating water micro-droplets over a heated solid substrate and suggest a model for the mechanisms of both droplet levitation and inter-droplet interaction. According to their model, the reflection of the Stefan flow (due to droplet evaporation) off the substrate is the mechanism of levitation, while the same Stefan flow also results in droplet repulsion preventing them from merging. They further apply their model to explain the levitation of droplets over a liquid surface and suggest the $h/R \sim R^{-3/2}$ dependency for the droplet radius $R$ vs. the height of its center above the surface, $h$. While there is no doubt that the experimentally observed phenomenon is of interest, here we show that the observations should be interpreted differently.

Levitating monolayer clusters of micro-droplets above a locally heated water layer is a relatively recently discovered phenomenon [2]. Condensed droplets with a typical size of 10 μm - 200 μm levitate at an equilibrium height, where their weight is equilibrated by the drag force of the ascending air-vapor jet. At the same time, droplets are dragged towards the center of the heated spot; however, they do not merge forming an ordered hexagonal (densest packed) pattern due to an aerodynamic repulsive pressure force from gas flow between the droplets [3].

Evaporation and convection cause an ascending vapor and air flow above the water layer on a copper substrate heated to above 80 ºC studied in Ref. [1]. The vapor concentration over the small (<1 mm) dry spot is lower than above water (so that droplets evaporate); however, there is a gas flow, which drags the droplets up, even over the dry spot. A particle in a shear boundary-layer flow is subject to the lift force. The direct evidence of that is observed in the supplementary video S3 attached to the letter by Zaitsev at al. [1]. The droplet height $h$ can be easily measured since both the droplet and its reflection from the substrate are seen (**Fig. 1a**). After evaporating droplets reach a critical radius of about 5 μm, they fly up. Thus, for a droplet of a constant radius, the $h/R$ ratio increases from $h/R$=4.3 to $h/R$=19.8 (**Fig. 1b**). This is inconsistent with Eqs. 3-4 of Ref [1] and with the hypothesis that the Stefan flow reflected from the substrate is the major driving force of the levitation. Therefore, the levitation mechanism over the small dry island is the same as that over the surrounding water layer: an ascending gas jet. The only difference is lower vapor concentration over the dry area, which causes droplets to evaporate and shrink.

Furthermore, the mechanism considered in [1] as the main one cannot explain droplet levitation over a liquid surface and the long-range repulsion force. Unlike the evaporating droplets over the solid, droplets over the liquid layer grow due to condensation. Therefore, the Stefan flow cannot explain the repulsion between the droplets, and the contribution of the phase changes is negligible in comparison with that of ordinary gas flow. Accurate experiments show that growing droplets can sit very close to the liquid surface so that the height of the center increases with increasing radius, which is inconsistent with the proposed $h/R \sim (R-R_\infty)^{-1/2}$ trend (see, for example, Ref. [4] Fig. 2).

We therefore conclude that the Stefan flow is not the main mechanism of droplet interaction and levitation.

A. A. Fedorets[1], L. A. Dombrovsky[2], and M. Nosonovsky[1,3]

[1]Tyumen State University, Tyumen, Russia
[2]Joint Institute for High Temperatures, Moscow, Russia
[3]University of Wisconsin-Milwaukee, Milwaukee, WI 53201, USA

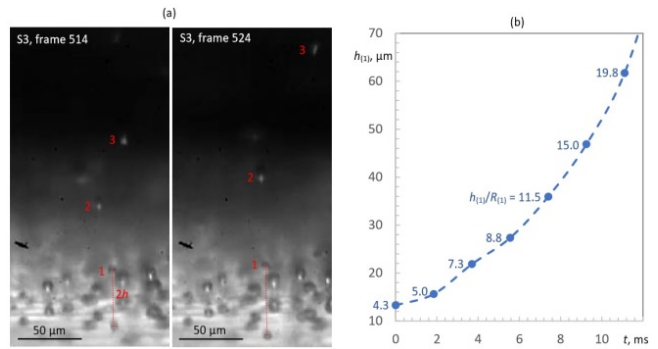

FIG. 1. (a) Frames of the video S3 [1] showing ascending droplets and (b) height of the droplet 1 as a function of time, from frame 494 to 554.